\documentclass[12pt]{article}
\usepackage{amsmath}
\usepackage{prodint}
\usepackage{graphicx}
\usepackage{enumerate}
\usepackage{natbib}
\usepackage{url} 
\usepackage{binomexp}
\usepackage{mathtools}
\usepackage{amsfonts}
\usepackage{mathrsfs}
\RequirePackage{subcaption}
\usepackage{booktabs}
\usepackage{bbm}
\usepackage{dsfont}
\usepackage{longtable}
\usepackage{yhmath}
\usepackage{bm}
\usepackage{amssymb}
\usepackage[nodisplayskipstretch]{setspace}
\usepackage{tikz}
\usepackage{float}
\usepackage{color}
\usepackage{stmaryrd}
\usepackage{paralist}
\usepackage{multirow}
\usepackage{threeparttable}
\usepackage{soul}
\usepackage{algorithm,algpseudocode}
\makeatletter
\newcommand{\algmargin}{\the\ALG@thistlm}
\makeatother
\newlength{\whilewidth}
\algdef{SE}[parWHILE]{parWhile}{EndparWhile}[1]
  {\parbox[t]{\dimexpr\linewidth-\algmargin}{%
     \hangindent\whilewidth\strut\algorithmicwhile\ #1\ \algorithmicdo\strut}}{\algorithmicend\ \algorithmicwhile}%
\algnewcommand{\parState}[1]{\State%
  \parbox[t]{\dimexpr\linewidth-\algmargin}{\strut #1\strut}}

\newcommand{\bfo}{\bm{o}}

\newcommand{\bfx}{\bm{x}}

\newcommand{\bfz}{\bm{z}}

\newcommand{\bfalpha}{\bm{\alpha}}

\newcommand{\bfSigma}{\bm{\Sigma}}

\newcommand{\rmd}{\mathrm{d}}

\newcommand{\Var}{{\mathbb{V}\mathrm{ar}}}
\newcommand{\Cov}{{\mathbb{C}\mathrm{ov}}}

\newcommand{\est}{{\mathrm{est}}}
\newcommand{\rr}{{\mathrm{rr}}}
\newcommand{\km}{{\mathrm{km}}}
\newcommand{\ipcw}{{\mathrm{ipcw}}}
\newcommand{\dml}{{\mathrm{dml}}}
\newcommand{\eff}{{\mathrm{eff}}}

\newcommand{\bfB}{\bm{B}}

\newcommand{\bfI}{\bm{I}}

\newcommand{\bfL}{\bm{L}}

\newcommand{\bfU}{\bm{U}}

\newcommand{\bfZ}{\bm{Z}}

\newcommand{\bbE}{\mathbb{E}}
\newcommand{\bbG}{\mathbb{G}}

\newcommand{\bbP}{\mathbb{P}}
\newcommand{\bbR}{\mathbb{R}}

\newcommand{\bbone}{\mathbbm{1}}

\newcommand{\calB}{\mathcal{B}}
\newcommand{\calC}{\mathcal{C}}
\newcommand{\calD}{\mathcal{D}}
\newcommand{\calE}{\mathcal{E}}

\newcommand{\calI}{\mathcal{I}}

\newcommand{\calN}{\mathcal{N}}
\newcommand{\calO}{\mathcal{O}}

\newcommand{\calT}{\mathcal{T}}
\newcommand{\calU}{\mathcal{U}}

\newcommand{\frakS}{\mathfrak{S}}

\newcommand{\frakZ}{\mathfrak{Z}}

\usepackage[colorlinks=true,linkcolor=red,citecolor=black]{hyperref}

\newtheorem{theorem}{Theorem}
\newtheorem{assumption}{Assumption}

\newcommand{\blind}{1}

\allowdisplaybreaks

\usetikzlibrary{shapes,decorations,arrows,calc,arrows.meta,fit,positioning}
\tikzset{
     -Latex,auto,node distance =1 cm and 1 cm,semithick,
     state/.style ={ellipse, draw, minimum width = 0.7 cm},
     point/.style = {circle, draw, inner sep=0.04cm,fill,node contents={}},
    bidirected/.style={Latex-Latex,dashed},
     el/.style = {inner sep=2pt, align=left, sloped}
 }

\begin{document}

\def\spacingset#1{\renewcommand{\baselinestretch}%
{#1}\small\normalsize} \spacingset{1}


\if1\blind
{
  \title{\Large \bf Asymptotic theory of rerandomization for survival analysis}
  \author{Xinyuan Chen$^{1,\ast}$ and Fan Li$^{2,\dagger}$\vspace{0.2cm}\\
    $^1$Department of Mathematics and Statistics,\\ Mississippi State University, MS, USA\\
    $^2$Department of Biostatistics, Yale School of Public Health, CT, USA\\
    ${}^\ast$xchen@math.msstate.edu\\
    ${}^\dagger$fan.f.li@yale.edu}  
  \maketitle
} \fi

\if0\blind
{
  \bigskip
  \bigskip
  \bigskip
  \begin{center}
    {\Large \bf Asymptotic theory of rerandomization for survival analysis}
\end{center}
  \medskip
} \fi

\bigskip
\begin{abstract}
	Rerandomization systematically reduces chance imbalance and can improve the efficiency of the average treatment effect estimator in randomized experiments. While the asymptotic properties of finite-dimensional M-estimators under rerandomization have been established, existing theory does not directly address survival outcomes under censoring, where the target estimand involves infinite-dimensional functional parameters. This article establishes the uniform weak convergence of treatment-specific survival function estimators under rerandomization and stratified rerandomization. We prove that the Kaplan-Meier and inverse probability of censoring weighted Kaplan-Meier estimators converge to tight limiting processes with reduced pointwise asymptotic variances. Furthermore, we prove that the pointwise asymptotic variance of the debiased machine learning survival function estimator remains invariant under rerandomization, a consequence of the Neyman orthogonality. Simulations and a real data example are used to illustrate the theoretical results. Our results characterize the geometric interplay between restricted randomization designs and analysis-stage covariate adjustment for functional target estimands in survival analysis.
\end{abstract}

\noindent%
{\it Keywords:} Bahadur representation; Covariate adjustment; Debiased machine learning; Inverse probability of censoring weighting; Randomized experiments; Rerandomization

\spacingset{1.75} 

\section{Introduction} \label{sec:intro}

Rerandomization \citep{Morgan2012} is an attractive experimental design strategy that restricts randomization to allocations that satisfy pre-specified covariate balance criteria, thereby improving baseline balance and potentially improving the statistical efficiency for inference. While the majority of the theoretical literature operates within the finite-population design-based framework \citep[e.g.,][]{Li2018PNAS, Li2020, li2020-2k, Wang2022, Yang2023, Wang2025, Han2026}, recent work by \citet{wang2024asymptotic} offered an important generalization of the rerandomization theory to super-population estimands and a wider class of estimators, by characterizing the asymptotic behavior of finite-dimensional M-estimators for scalar average causal effect estimands under a sampling-based perspective. Furthermore, \citet{Wang2023stratify} studied the stratified rerandomization with a fixed number of strata, and \citet{bai2023efficiency} and \citet{cytrynbaum2024finely} considered highly and finely stratified rerandomization designs, establishing efficiency and asymptotic theory for M-estimators of finite-dimensional treatment effect parameters. A comprehensive review of historical foundations and theoretical developments of rerandomization, with a focus on the average treatment effect estimand, is provided in \citet{junior2025does}. 

Beyond simple scalar estimands, causal survival analysis has been central in randomized experiments, where time-to-event outcomes such as overall and progression-free survival serve as primary endpoints in many pivotal trials. The existing asymptotic theory for rerandomization, however, does not directly apply to this setting. More importantly, causal survival analysis inherently targets infinite-dimensional functional parameters---most typically the treatment-specific/counterfactual survival function \citep{fay2024causal}---and is subject to right-censoring and time-evolving risk sets. Characterizing the impact of rerandomization on causal survival analysis thus requires advancing the technical machinery from standard multivariate central limit theorems to the uniform weak convergence of empirical processes. Motivated by this methodological gap, we seek to establish the variance-reduction geometry and uniform limiting processes for estimating treatment-specific survival functions under rerandomization.

We investigate the asymptotic implications of rerandomization on three causal survival estimators: the Kaplan-Meier (KM) estimator, the inverse probability of censoring weighted Kaplan-Meier (IPCW-KM) estimator, and the recently established debiased machine learning (DML) estimator \citep{Westling2024}, which leverages the efficient influence function (EIF) to achieve flexible covariate adjustment in the analysis stage. Our first finding is that under rerandomization, all three estimators remain uniformly consistent and preserve their Bahadur representations under simple randomization. However, under rerandomization, the KM and IPCW-KM estimators converge uniformly to tight limiting processes with reduced pointwise asymptotic variances unless the outcome counting and at-risk processes are uncorrelated with the rerandomization covariates. In contrast, rerandomization does not reduce the pointwise asymptotic variance of the DML estimator, provided the rerandomization covariates are included in the set of covariates adjusted for in the nuisance outcome model; hence, rerandomization is asymptotically ignorable in the analysis stage with flexible covariate adjustment. These results, for the first time, expand the asymptotic analysis of \citet{wang2024asymptotic} and the insights under rerandomization to the infinite-dimensional functional domain. 

To understand why rerandomization reduces the variance of the KM and IPCW-KM estimators but not the DML estimator, we appeal to a geometric decomposition. By the Hilbert space projection theorem, each estimator's limiting process decomposes into a projection onto the subspace spanned by the rerandomization covariates and an independent Gaussian process residual orthogonal to that subspace. Hence, the randomness of the projection onto the rerandomization covariate subspace arises solely from the covariates, with the projection processes being deterministic. The rerandomization constraint restricts the stochastic variability of this projection component, whereas the residual component remains unaffected, effectively clamping the stochastic process cloud into a tighter spindle. For the KM and IPCW-KM estimators, this projection is non-trivial because the outcome counting and at-risk processes are generally correlated with the rerandomization covariates. In particular, although the IPCW-KM estimator adjusts for covariates to account for censoring under the covariate-dependent censoring assumption, the outcome counting and at-risk processes remain unadjusted, so the variance-reduction implication of rerandomization for the KM estimator applies equally even in the presence of censoring weights. For the DML estimator, however, the Neyman orthogonality of the EIF renders the projection asymptotically zero, leaving no variance to clamp. We formalize this geometric picture and its implications for both pointwise and uniform inference of the three typical causal survival estimators.

In developing the new asymptotic theory when introducing rerandomization to causal survival analysis, we assume that the true survival functions of the outcome and censoring times are c\`adl\`ag (right continuous with left limits) to allow greater generality. The rest of the article is organized as follows. Section \ref{sec:preliminaries} introduces the preliminaries, including structural assumptions, infinite-dimensional causal estimands, and the rerandomization design. We also review the construction of these three estimators with their existing properties under simple randomization. Section \ref{sec:asymptotics} gives the asymptotic properties of estimators under rerandomization, and Section \ref{sec:stratified} further extends our new asymptotic results to stratified rerandomization. Section \ref{sec:simulation} presents the numerical studies to illustrate the asymptotic analysis under rerandomization, including simulations and the analysis of a real-life dataset. Section \ref{sec:conclusion} concludes with final remarks. Proofs of theoretical results and additional simulation results are available in the Supplementary Materials.

\section{Preliminaries} \label{sec:preliminaries}

\subsection{Assumptions and survival function estimands}

We consider a randomized experiment with $n$ units. Let $A_i=a\in\{0,1\}$ denote the treatment assignment for unit $i=1,\ldots,n$. Under the potential outcome framework, we write the potential outcome and censoring time for unit $i$ under treatment $a$ as $T_i(a)$ and $C_i(a)$, respectively. The baseline covariates are written as $\bfZ_i\in\frakZ\subset\bbR^p$. The complete but not fully observed data vector for each individual is $\calC_i=\{T_i(1),T_i(0),C_i(1),C_i(0),\bfZ_i\}$.

\begin{assumption}[Super-population] \label{asp:super-population}
	The complete data vectors $(\calC_1,\ldots,\calC_n)$ are independent and identically distributed draws from an unknown distribution with finite second moments. 
\end{assumption}

Assumption \ref{asp:super-population} postulates a notional super-population from which the observed sample is randomly drawn, and defines the target estimands as features of this super-population distribution. This is not a limitation but rather a natural modeling choice. That is, the sampling-based framework is by far the most common inferential paradigm for survival analysis, underpinning virtually all standard developments and textbooks \citep[e.g.,][]{fleming2011counting}. Adopting this framework facilitates direct comparison with existing asymptotic results for survival function estimators established under simple randomization. Next, we let $S_{T,a}(t)=\bbP\{T(a)>t\}$, $S_{C,a}(t)=\bbP\{C(a)>t\}$, $S_{T,a}(t|\bfz)=\bbP\{T(a)>t|\bfZ=\bfz\}$, $S_{C,a}(t|\bfz)=\bbP\{C(a)>t|\bfZ=\bfz\}$ for $\bfz\in\frakZ$ denote the marginal and conditional survival functions of $T(a)$ and $C(a)$. 

\begin{assumption}[Censoring mechanism] \label{asp:censor}
	(i) $\{T(1),T(0)\}\perp\{C(1),C(0)\}$. (ii) $\{T(1), T(0)\}\perp\{C(1),C(0)\}|\bfZ$. (iii) For $a=0,1$, the maximum follow-up $\tau\in(0,\infty)$ with $S_{T,a}(\tau)>0$ and $S_{C,a}(\tau)>0$, as well as $S_{T,a}(\tau|\bfz)>0$ and $S_{C,a}(\tau|\bfz)>0$ for $\bbP$-almost all $\bfz \in \frakZ$.
\end{assumption}

Assumption \ref{asp:censor}(i) and (ii) state the two considered censoring settings, i.e., the completely independent censoring and the covariate-dependent censoring. Under simple randomization, the KM estimator is justified in the former setting, whereas the IPCW-KM estimator is required in the latter for consistent estimation (provided that the censoring weights are correctly specified). The DML estimator, on the other hand, is suitable in both settings. Assumption \ref{asp:censor}(iii) is a regularity condition requiring finite maximal follow-up and (conditional) survival functions being bounded away from zero at the maximal follow-up to prevent ill-behaviors of the estimators. Assumption \ref{asp:censor} is a typical condition invoked for survival analysis with right-censored time-to-event outcomes. 

Our target estimand is the treatment-specific survival function $S_{T,a}(t)$ for $t\in[0,\tau]$ and $a=0,1$, which can further be used in forming other summary estimands, e.g., the survival probability causal effect at time $t$, $\Delta S_T(t)\equiv S_{T,1}(t)- S_{T,0}(t)$ \citep{fay2024causal}. The commonly used summary estimands are oftentimes known regular functionals of $S_{T,a}(t)$, and the theoretical properties of their corresponding estimators can be obtained straightforwardly via the functional Delta method \citep[\S20]{vandervaart1998}. Therefore, we subsequently focus our discussions on the estimation of $S_{T,a}(t)$ as this is the essential building block.

We also define $\Lambda_{T,a}(t)$ and $\Lambda_{C,a}(t)$ as the cumulative hazard of $T(a)$ and $C(a)$, respectively, and similarly, $\Lambda_{T,a}(t|\bfz)$ and $\Lambda_{C,a}(t|\bfz)$ are the conditional cumulative hazards. With the cumulative hazard functions, the survival functions can be written as
\begin{align*} 
	S(t) = \Prodi_{[0,t]}\left\{1-\rmd\Lambda(s)\right\}, \quad S(t|\bfz) = \Prodi_{[0,t]}\left\{1-\rmd\Lambda(s|\bfz)\right\},
\end{align*}
for $t\in[0,\tau]$, where $\Prodi_{[0,t]}\rmd f(s)$ is the product integral of function $f$ over $[0,t]$ \citep{Gill1990}, with $(S,\Lambda)=(S_{T,a},\Lambda_{T,a})$ or $(S_{C,a},\Lambda_{C,a})$. Note that Assumption \ref{asp:censor}(iii) serves as a regularity condition to rule out the scenario where the jump $\rmd\Lambda=1$ for $t\in[0,\tau]$ by requiring the (conditional) survival functions to be bounded away from zero at the maximum follow-up time $\tau$. 

\subsection{Simple randomization and rerandomization}

Let $A^*=a\in\{0,1\}$ denote the treatment assignment under simple randomization, where $\bbP(A^*=a)=\pi_a\in(0,1)$ with $A^*\perp\calC$. Compared to simple randomization, rerandomization controls imbalance on a subset of $p^\rr$ baseline covariates, $\bfZ^\rr\subseteq\bfZ$ \citep{Morgan2012}, which involves the following steps. First, one independently generates $(A_1^*,\ldots,A_n^*)$ from a Bernoulli distribution with $\bbP(A_i^*=a)=\pi_a\in(0,1)$ under simple randomization. Then, the imbalance statistic on $(\bfZ_1^\rr,\ldots,\bfZ_n^\rr)$ and its variance estimator are computed as
\begin{align*}
	&\bfI_n^* = \frac{1}{n_1^*}\sum_{i=1}^n A_i^*\bfZ_i^\rr - \frac{1}{n_0^*}\sum_{i=1}^n (1-A_i^*)\bfZ_i^\rr,\displaybreak[0]\\
	&\widehat{\Var}(\bfI_n^*) = \frac{1}{n_1^*n_0^*}\sum_{i=1}^n\left(\bfZ_i^\rr-\overline\bfZ^\rr\right)\left(\bfZ_i^\rr-\overline\bfZ^\rr\right)^\top,
\end{align*}
where $n_a^*=\sum_{i=1}^n\bbone(A_i^*=a)$ and $\overline\bfZ^\rr=n^{-1}\sum_{i=1}^n\bfZ_i^\rr$. Lastly, given a pre-specified balance threshold $0<c<\infty$, one checks whether the realized assignment satisfies the balance criterion with $\bfI_n^{\ast\top}\{\widehat{\Var}(\bfI_n^*)\}^{-1}\bfI_n^*<c$. If true, the final treatment assignment is $(A_1,\ldots,A_n)=(A_1^*,\ldots,A_n^*)$; otherwise, the steps are repeated until the first randomization scheme satisfying the balance criterion. It is worth mentioning that the above procedure considers the Mahalanobis distance to define the balance criterion, and thus $\bfI_n^{\ast\top}\{\widehat{\Var}(\bfI_n^*)\}^{-1}\bfI_n^*$ follows a chi-squared distribution asymptotically. A key analytic implication is that rerandomization restricts the space of acceptable assignments to those satisfying $\bfI_n^{\ast\top}\{\widehat{\Var}(\bfI_n^*)\}^{-1}\bfI_n^*<c$, thereby inducing correlation among the treatment assignments $(A_1,\ldots,A_n)$.

\begin{assumption}[Rerandomization] \label{asp:rerandomization}
	(i) The covariance matrix of the imbalance criterion $\widehat{\Var}(\bfI_n^*)$ is positive definite, and $n\widehat{\Var}(\bfI_n^*)\overset{p}{\to}(\pi_1\pi_0)^{-1}\bfSigma_{\bfZ^\rr}$, where $\bfSigma_{\bfZ^\rr}\equiv\Cov(\bfZ_i^\rr)$, which is also positive definite. (ii) Let $\calE_n=\{\bfI_n^{\ast\top}\{\widehat{\Var}(\bfI_n^*)\}^{-1}\bfI_n^*<c\}$ denote the acceptance set for rerandomization. Then, $\calE=\lim_{n\to\infty}\calE_n$ with $\bbP(\partial\calE)=0$, where $\partial\calE$ is the boundary of $\calE$.
\end{assumption}

Assumption \ref{asp:rerandomization} ensures that the acceptance set $\calE_n$ and its limit $\calE$ are not degenerate and there are no discontinuities on the boundary of $\calE$. Conditional on $(\calC_1,\ldots,\calC_n)$, the estimated covariance matrix of the imbalance criterion is     
\begin{align*}
	\widehat\Var(\bfI_n^*) = \left( \frac{1}{n_a^*} + \frac{1}{n_{1-a}^*} \right) \frac{1}{n-1} \sum_{i=1}^n \left(\bfZ_i^\rr - \overline\bfZ_i^\rr\right)\left(\bfZ_i^\rr - \overline\bfZ_i^\rr\right)^\top \equiv \left( \frac{1}{n_a^*} + \frac{1}{n_{1-a}^*} \right) \bfSigma_{\bfZ^\rr,n}.
\end{align*}
To prepare for the subsequent asymptotic analyses, we further define 
\begin{align} \label{eq:B-star}
	\bfB_n^* = \frac{1}{\sqrt n}\sum_{i=1}^n\frac{1}{\pi_1\pi_0}(A_i-\pi_1)\left\{\bfZ_i^\rr-\bbE(\bfZ_i^\rr)\right\},
\end{align}
and note that $\sqrt n\bfI_n^* = \bfB_n^* + o_{\bbP}(1)$. Lemma S1 in Section S2.1 of the Supplementary Materials shows that $\calE_n = \{ \bfB_n^{*\top} \bfSigma_{\bfB^*,n}^{-1} \bfB_n^* < c \} + o_{\bbP}(1)$, where $\bfSigma_{\bfB^*,n} = (\pi_1\pi_0)^{-1}\bfSigma_{\bfZ^\rr,n}$. Therefore, the limiting acceptance set for rerandomization can be written as $\calE=\{\bfx\in\bbR^{p^\rr}:\bfx^\top\bfSigma_{\bfB^*}^{-1}\bfx < c\}$, which is a convex ellipsoid, where $\bfSigma_{\bfB^*} = (\pi_1\pi_0)^{-1}\bfSigma_{\bfZ^\rr}$.

\subsection{Three typical survival estimators and their properties under simple randomization} \label{sec:DML}

Under the stable treatment and value assumption (SUTVA), we have $T_i=A_iT_i(1)+(1-A_i)T_i(0)$ and $C_i=A_iC_i(1)+(1-A_i)C_i(0)$, as well as the observed outcome time $X_i\equiv T_i\wedge C_i$ and censoring indicator $\Delta_i\equiv\bbone(T_i\leq C_i)$. For notational purposes, we define the outcome and censoring counting processes as $N_{T,i}(t)=\bbone(X_i\leq t,\Delta_i=1)$ and $N_{C,i}(t)=\bbone(X_i\leq t,\Delta_i=0)$, and the at-risk process as $Y_i(t)=\bbone(X_i\geq t)$. The observed data vector is written as $\calO_i=(A_i,X_i,\Delta_i,\bfZ_i)$ with $\calO_i(a)=\{a,X_i(a),\Delta_i(a),\bfZ_i\}$, where $X_i(a)\equiv T_i(a)\wedge C_i(a)$ and $\Delta_i(a)\equiv\bbone\{T_i(a)\leq C_i(a)\}$. We study the asymptotic properties of the following three estimators for the treatment-specific survival function estimand. 

\textbf{KM and IPCW-KM estimators.} Under completely independent censoring, Assumption \ref{asp:censor}(i) and (iii), we can estimate $\Lambda_{T,a}(t)$ via the Nelson-Aalen (NA) estimator, and $S_{T,a}(t)$ via the KM estimator, which are given by
\begin{align*}
	\widehat\Lambda_{T,a}^\km(t) = \frac{\sum_{i=1}^n\bbone(A_i=a)\rmd N_{T,i}(s)}{\sum_{i=1}^n\bbone(A_i=a)Y_i(s)}, \quad \widehat S_{T,a}^\km(t) = \Prodi_{[0,t]}\left\{1-\rmd\widehat\Lambda_{T,a}^\km(s)\right\},
\end{align*}
respectively, for $t\in[0,\tau]$. Under simple randomization, it is well-established that $\widehat S_{T,a}^\km(t)$ is uniformly consistent, admits a Bahadur representation with influence function $\varphi_{T,a}^\km(t,\calO)$ uniformly over $t\in[0,\tau]$, and the process $\{\sqrt n\{\widehat S_{T,a}^\km(t)-S_{T,a}(t)\}:t\in[0,\tau]\}$ converges weakly to a mean-zero Gaussian process $\bbG_{T,a}^\km(\cdot)$ on $[0,\tau]$ with covariance function $\Sigma_{T,a}^\km(s,t)=\bbE\{\varphi_{T,a}^\km(s,\calO)\varphi_{T,a}^\km(t,\calO)\}$ \citep{Wang1987}. In the possibly more realistic, covariate-dependent censoring setting, under Assumption \ref{asp:censor}(ii) and (iii), we can consistently estimate $\Lambda_{T,a}(t)$ via the IPCW-NA estimator, and $S_{T,a}(t)$ via the IPCW-KM estimator, which are
\begin{align*}
	\widehat \Lambda_{T,a}^\ipcw(t) = \frac{\sum_{i=1}^n\widehat w_a(s,\bfZ_i)\rmd N_{T,i}(s)}{\sum_{i=1}^n\widehat w_a(s,\bfZ_i)Y_i(s)}, \quad \widehat S_{T,a}^\ipcw(t) = \Prodi_{[0,t]}\left\{1-\rmd\widehat\Lambda_{T,a}^\ipcw(s)\right\},
\end{align*}
where $\widehat w_a(s,\bfZ_i)=\bbone(A_i=a)/\widehat S_{C,a}(s-|\bfZ_i)$. When the censoring model is correctly specified (e.g., by Cox proportional hazards regression), under simple randomization, $\widehat S_{T,a}^\ipcw(t)$ is uniformly consistent, admits a Bahadur representation with influence function $\varphi_{T,a}^\ipcw(t,\calO)$ uniformly over $t\in[0,\tau]$, and $\{\sqrt n\{\widehat S_{T,a}^\ipcw(t)-S_{T,a}(t)\}:t\in[0,\tau]\}$ converges weakly to a mean-zero Gaussian process $\bbG_{T,a}^\ipcw(\cdot)$ on $[0,\tau]$ with covariance function $\Sigma_{T,a}^\ipcw(s,t)=\bbE\{\varphi_{T,a}^\ipcw(s,\calO)\varphi_{T,a}^\ipcw(t,\calO)\}$ \citep{vanderlaan2003}. In particular, compared to the influence function $\varphi_{T,a}^\km(t,\calO)$, $\varphi_{T,a}^\ipcw(t,\calO)$ and the limiting covariance process $\Sigma_{T,a}^\ipcw(s,t)=\bbE\{\varphi_{T,a}^\ipcw(s,\calO)\varphi_{T,a}^\ipcw(t,\calO)\}$ include additional terms arising from modeling the censoring process.

\textbf{DML estimator.} Beyond the KM and IPCW-KM estimators considered above, data-adaptive machine learning methods offer additional flexibility to leverage baseline covariate information and can achieve the semiparametric efficiency lower bound in randomized experiments \citep{Chernozhukov2018}. In the survival analysis context, the appeal of such an efficient estimator can be salient, as covariate adjustment serves a dual purpose; that is, adjusting for covariates in the censoring model via IPCW corrects for the selection bias induced by right censoring, while simultaneously adjusting for covariates in the survival outcome model improves estimation precision. The DML estimator, constructed from the EIF with cross-fitted nuisance estimates, incorporates both sources of covariate information and attains the optimal asymptotic variance under simple randomization \citep{Westling2024}. However, the validity and efficiency optimality of such estimators under rerandomization remain unexplored for survival outcomes. We next introduce the DML estimator for the survival function and subsequently characterize its asymptotic behavior under rerandomization.

For the DML estimator, we primarily consider the covariate-dependent censoring mechanism for brevity, casting the completely independent censoring mechanism as a special case. Specifically, under Assumption \ref{asp:censor}(ii) and (iii), \citet{Westling2024} showed that the EIF of $S_{T,a}(t)$ is $\varphi_{T,a}^\eff(t,\calO)=\varphi_{T,a}^{\eff,\ast}(t,\calO)- S_{T,a}(t)$, where
\begin{equation}
	\begin{aligned} \label{eq:EIF}
		\varphi_{T,a}^{\eff,\ast}(t,\calO)= S_{T,a}(t|\bfZ)\Bigg[1-\frac{\bbone(A=a)}{\pi_a}&\Bigg\{\frac{\bbone(X\leq t,\Delta=1)}{S_{T,a}(X-|\bfZ)S_{C,a}(X-|\bfZ)}\\
		&\quad - \int_0^{t\wedge X}\frac{\rmd \Lambda_{T,a}(s|\bfZ)}{S_{T,a}(s-|\bfZ)S_{C,a}(s-|\bfZ)}\Bigg\}\Bigg].
	\end{aligned}
\end{equation}

Expanding the results in \citet{Chernozhukov2018}, they developed the cross-fitted, EIF-based DML estimator $\widehat S_{T,a}^\dml(t)=\bbP_{n,K}\widehat\varphi_{T,a}^{\eff,\ast}(t,\calO)$, where $\widehat\varphi_{T,a}^{\eff,\ast}(t,\calO)$ is $\varphi_{T,a}^{\eff,\ast}(t,\calO)$ in \eqref{eq:EIF} with the nuisance functions $\xi(t|\bfz)\equiv\{S_{T,a}(t|\bfz),\allowbreak S_{C,a}(t|\bfz), \Lambda_{T,a}(t|\bfz)\}$ replaced by their respective estimates. For a deterministic integer $K$, the cross-fitting scheme randomly partitions the indices $\calI=\{1, 2,\ldots,n\}$ into $K$ disjoint sets $\calI_1, \calI_2, \ldots, \calI_K$ with $||\calI_k|-K^{-1}n|\leq1$ for $k=1,\ldots,K$. For each $k$, the nuisance functions $\xi(t|\bfz)$ are estimated on $\calI_k^c=\calI\setminus\calI_k$ to obtain estimates $\widehat\xi_k(t|\bfz)\equiv\{\widehat S_{T,a,k}(t|\bfz), \widehat S_{C,a,k}(t|\bfz), \widehat\Lambda_{T,a,k}(t|\bfz)\}$ and $\widehat\varphi_{T,a,k}^{\eff,\ast}(t,\calO_i)$ is formed on $\calI_k$ with estimated nuisances being $\widehat\xi_k(t|\bfz)$. Therefore,
\begin{align*}
	\bbP_{n,K}\widehat\varphi_{T,a}^{\eff,\ast}(t,\calO)=\frac{1}{n}\sum_{k=1}^K\sum_{i\in\calI_k}\widehat\varphi_{T,a,k}^{\eff,\ast}(t,\calO_i).
\end{align*}

\citet{Westling2024} proved that $\widehat S_{T,a}^\dml(t)$ is uniformly consistent, admits a Bahadur representation with influence function $\varphi_{T,a}^\eff(t,\calO)$ uniformly over $t\in[0,\tau]$, and $\{\sqrt n\{\widehat S_{T,a}^\dml(t)-S_{T,a}(t)\}:t\in[0,\tau]\}$ converges weakly to a mean-zero Gaussian process $\bbG_{T,a}^\dml(\cdot)$ on $[0,\tau]$ with covariance function $\Sigma_{T,a}^\dml(s,t)=\bbE\{\varphi_{T,a}^\eff(s,\calO)\varphi_{T,a}^\eff(t,\calO)\}$.

\section{Asymptotic survival analysis under rerandomization} \label{sec:asymptotics}

\subsection{KM and IPCW-KM estimators}

We first introduce regularity conditions for establishing the uniform consistency of $\widehat S_{T,a}^\km(t)$ and $\widehat S_{T,a}^\ipcw(t)$ for $t\in[0,\tau]$ under rerandomization.
\begin{itemize}
	\item[(C1)] $S_{T,a}(t)$ and $S_{C,a}(t)$ are c\`adl\`ag functions on $[0,\tau]$. For $\bbP$-almost all $\bfz\in\frakZ$, $S_{T,a}(t|\bfz)$ and $S_{C,a}(t|\bfz)$ are c\`adl\`ag functions on $[0,\tau]$.
	\item[(C2)] Let $\widehat S_{C,a}(t|\bfz)$ be an estimator for $S_{C,a}(t|\bfz)$. Under simple randomization, for $\bbP$-almost all $\bfz\in\frakZ$, $\sup_{t\in[0,\tau]}|\widehat S_{C,a}(t|\bfz)- S_{C,a}(t|\bfz)| \overset{p}{\to} 0$, and there exists $\epsilon>0$ such that $\inf_{t\in[0,\tau]}S_{C,a}(t|\bfz)\geq \epsilon$ and $\inf_{t\in[0,\tau]}\widehat S_{C,a}(t|\bfz)\geq \epsilon/2$.
\end{itemize}

Condition (C1) states that the marginal survival functions and conditional survival functions for $\bbP$-almost all $\bfz\in\frakZ$ are c\`adl\`ag on $[0,\tau]$, which goes beyond the class of continuous survival functions for greater generality. Condition (C2) involves two components. First, it assumes the existence of a uniformly consistent estimator $\widehat S_{C,a}(t|\bfz)$ for the conditional censoring survival function $S_{C,a}(t|\bfz)$. This is a modeling assumption on the censoring mechanism. It essentially requires that the analyst posits a correctly specified regression model for the censoring process conditional on baseline covariates, such as a parametric survival model or a Cox proportional hazards regression, both of which yield uniformly consistent estimators under standard regularity conditions (assuming the specified censoring model matches the underlying data-generating process). Second, it requires that both $S_{C,a}(t|\bfz)$ and $\widehat S_{C,a}(t|\bfz)$ are bounded away from zero on $[0,\tau]$ for $\bbP$-almost all $\bfz\in\frakZ$, which ensures that the inverse probability of censoring weights remain bounded and prevents extreme weights from dominating the IPCW-KM estimator. Together, these two components formalize the requirements for a well-behaved censoring model that underpins valid IPCW estimation.

\begin{theorem} \label{thm:KM-IPCW-consistency}
	(i) Under the rerandomization design, Assumptions \ref{asp:super-population}, \ref{asp:censor}(i) and (iii), and \ref{asp:rerandomization}, and Condition (C1), $\sup_{t\in[0,\tau]}|\widehat S_{T,a}^\km(t)-S_{T,a}(t)|\overset{p}{\to}0$. (ii) Under the rerandomization design, Assumptions \ref{asp:super-population}, \ref{asp:censor}(ii) and (iii), and Conditions (C1) and (C2), $\sup_{t\in[0,\tau]}|\widehat S_{T,a}^\ipcw(t)-S_{T,a}(t)|\overset{p}{\to}0$.
\end{theorem}

Theorem \ref{thm:KM-IPCW-consistency} shows that the uniform consistency of $\widehat S_{T,a}^\km(t)$ and $\widehat S_{T,a}^\ipcw(t)$ for $t\in[0,\tau]$ remain unchanged under rerandomization. This result is intuitive, as rerandomization only changes the treatment assignments and does not alter the within-treatment outcome mechanisms once the treatment assignments are determined. We then introduce regularity conditions for establishing the Bahadur representation and uniform weak convergence of $\widehat S_{T,a}^\km(t)$ and $\widehat S_{T,a}^\ipcw(t)$ for $t\in[0,\tau]$ under rerandomization.

\begin{itemize}
	\item[(A1)] Under simple randomization, for $\bbP$-almost all $\bfz\in\frakZ$,
	\begin{align*}
		\sqrt n\left\{\widehat S_{C,a}(t|\bfz)- S_{C,a}(t|\bfz)\right\} = \frac{1}{\sqrt n}\sum_{i=1}^n\varphi_{C,a}(t,\bfz,\calO_i) + o_{\bbP^*}(1),
	\end{align*}
	where $\widehat S_{C,a}(t|\bfz)$ is a parametric or semi-parametric estimator (e.g., Cox proportional hazards regression), $\varphi_{C,a}(t,\bfz,\calO_i)$ is the mean-zero influence function with finite pointwise variance for all $t\in[0,\tau]$ and $\bbP$-almost all $\bfz\in\frakZ$, the class $\{\varphi_{C,a}(t,\bfz,\bfo):t\in[0,\tau]\}$ is $\bbP$-Donsker, and $o_{\bbP^*}(1)$ represents a sequence of random variables converging to 0 in probability uniformly over $t\in[0,\tau]$.
\end{itemize}

Condition (A1) assumes that $\widehat S_{C,a}(t|\bfz)$ is a parametric or semi-parametric estimator that admits a Bahadur representation uniformly over $t\in[0,\tau]$, and the influence function of $\widehat S_{C,a}(t|\bfz)$ is in a $\bbP$-Donsker class indexed by $t$. These are commonly adopted regularity conditions for the IPCW-KM estimator \citep{vanderlaan2003}.

\begin{theorem} \label{thm:KM-IPCW-weak-convergence}
	Under the rerandomization design, Assumptions \ref{asp:super-population}-\ref{asp:rerandomization}, and Conditions (C1) and (A1), for $\est=\km,\ipcw$ and all $t\in[0,\tau]$, we have
	\begin{align} \label{eq:KM-Bahadur}
		\sqrt n\left\{\widehat S_{T,a}^\est(t)-S_{T,a}(t)\right\} = \frac{1}{\sqrt n}\sum_{i=1}^n\varphi_{T,a}^\est(t,\calO_i) + o_{\bbP^*}(1).
	\end{align}
	The process $\{\sqrt n\{\widehat S_{T,a}^\est(t)-S_{T,a}(t)\}:t\in[0,\tau]\}$ converges weakly to a mean-zero process 
	\begin{align} \label{eq:KM-limiting-process}
		\left\{V_a^\est(\cdot)\right\}^{1/2}  \left[\left\{\rho_a^\est(\cdot)\right\}^{1/2} \cdot r_a^\est(\cdot) + \left\{1 - \rho_a^\est(\cdot)\right\}^{1/2} \cdot \zeta_a^\est(\cdot)\right]
	\end{align}
	on $[0,\tau]$. Here, $r_a^\est(\cdot)$ and $\zeta_a^\est(\cdot)$ are two mean-zero, tight processes with $r_a^\est(\cdot)\perp \zeta_a^\est(\cdot)$, $V_a^\est(t)=\Sigma_{T,a}^\est(t,t)$, and 
	\begin{align} \label{eq:KM-variance-reduction}
		\rho_a^\est(t) = \frac{\bfSigma_{T,\bfB^*,a}^\est(t)^\top \bfSigma_{\bfB^*}^{-1} \bfSigma_{T,\bfB^*,a}^\est(t)}{\Sigma_{T,a}^\est(t,t)},
	\end{align}
	where
	\begin{align} \label{eq:KM-covariance}
		\bfSigma_{T,\bfB^*,a}^\est(t) = \bbE\left(\left[\varphi_{T,a}^\est\{t, \calO_i(a)\}-\varphi_{T,a}^\est\{t, \calO_i(1-a)\}\right]\left\{\bfZ_i^\rr - \bbE(\bfZ_i^\rr) \right\}\right).
	\end{align}
	In this representation, the process $r_a^\est(\cdot)=\bfU_a^\est(\cdot)^\top\bfL$, where $\bfU_a^\est(\cdot)=\bfalpha_a^\est(\cdot)/\|\bfalpha_a^\est(\cdot)\|$ with $\bfalpha_a^\est(\cdot)=\bfSigma_{\bfB^*}^{-1/2} \bfSigma_{T,\bfB^*,a}^\est(\cdot)$; for $t\in[0,\tau]$, $\Var\{r_a^\est(t)\} = \kappa(c) = \bbE(L_1^2 | \bfL^\top \bfL < c)$, with $L_1$ being the first element of $\bfL\sim\calN(\bm 0,\mathbf I_{p^\rr})$ conditioned on $\bfL^\top \bfL \leq c$, and for $s, t \in [0,\tau]$, $\Cov\{r_a^\est(s), r_a^\est(t)\}  = \kappa(c) \bfU_a^\est(s)^\top \bfU_a^\est(t)$; furthermore, the process $\zeta_a^\est(\cdot)$ is a standardized Gaussian process on $[0,\tau]$, with 
	\begin{align*}
		\Cov\left\{\zeta_a^\est(s),\zeta_a^\est(t)\right\} = \frac{\Sigma_{T,a}^\est(s,t)-\bfSigma_{T,\bfB^*,a}^\est(s)^\top \bfSigma_{\bfB^*}^{-1} \bfSigma_{T,\bfB^*,a}^\est(t)}{[V_a^\est(s)\{1 - \rho_a^\est(s)\}]^{1/2}[V_a^\est(t)\{1 - \rho_a^\est(t)\}]^{1/2}},
	\end{align*}
	for $s, t \in [0,\tau]$. 
\end{theorem}

Theorem \ref{thm:KM-IPCW-weak-convergence} proves the Bahadur representation and uniform weak convergence of the KM and IPCW-KM estimators under rerandomization over $t\in[0,\tau]$. The Bahadur representation is given in \eqref{eq:KM-Bahadur}, where the influence functions of the KM and IPCW-KM estimators remain identical regardless of simple randomization or rerandomization. Similar results for the KM estimator under the stratified and biased-coin randomization designs were shown in \citet{Wang2023}, and we prove a more general result under rerandomization. The limiting process in \eqref{eq:KM-limiting-process}, however, is different under rerandomization, which is obtained by first deriving the joint limiting process of $(n^{-1/2}\sum_{i=1}^n\varphi_{T,a}^\est(t,\calO_i),\bfB_n^*)$ with $\bfB_n^*$ given in \eqref{eq:B-star} and then conditioning on the event of $\lim_{n\to\infty}\bfB_n^*\in\calE$. Specifically, the first term in \eqref{eq:KM-limiting-process} is $\{V_a^\est(\cdot)\}^{1/2}\{\rho_a^\est(\cdot)\}^{1/2} \cdot r_a^\est(\cdot) = \bfSigma_{\bfZ^\rr}^{-1/2} \bfSigma_{T,\bfB^*,a}^\est(\cdot)^\top\bfL$, which is the projection of the estimator onto the subspace spanned by the rerandomization covariates, where the projection process $\bfSigma_{\bfB^*}^{-1/2} \bfSigma_{T,\bfB^*,a}^\est(\cdot)^\top$ is deterministic, and the randomness solely comes from $\bfL$, whose variation is restricted by the rerandomization constraint $\bfL^\top \bfL \leq c$. The second term in \eqref{eq:KM-limiting-process} is $\{V_a^\est(\cdot)\}^{1/2} \{1 - \rho_a^\est(\cdot)\}^{1/2} \cdot \zeta_a^\est(\cdot)$, which is the residual Gaussian process with covariance function $\Sigma_{T,a}^\est(s,t)-\bfSigma_{T,\bfB^*,a}^\est(s)^\top \bfSigma_{\bfB^*}^{-1} \bfSigma_{T,\bfB^*,a}^\est(t)$ and remains orthogonal to the projection and does not contain the balance threshold $c$ or variance multiplier $\kappa(c)$. Hence, it is unaffected by the rerandomization constraint.

Due to the orthogonality between the two terms in \eqref{eq:KM-limiting-process}, the pointwise variance of the limiting process can be obtained as the sum of the pointwise variances of these two terms. The pointwise variance of the first term in \eqref{eq:KM-limiting-process} is $\kappa(c)\bfSigma_{T,\bfB^*,a}^\est(t)^\top \bfSigma_{\bfB^*}^{-1} \bfSigma_{T,\bfB^*,a}^\est(t)$, and that of the second term is $\Sigma_{T,a}^\est(t,t)-\bfSigma_{T,\bfB^*,a}^\est(t)^\top \bfSigma_{\bfB^*}^{-1} \bfSigma_{T,\bfB^*,a}^\est(t)$, which leads to the total variance as $\Sigma_{T,a}^\est(t,t)-\{1-\kappa(c)\}\bfSigma_{T,\bfB^*,a}^\est(t)^\top \bfSigma_{\bfB^*}^{-1} \bfSigma_{T,\bfB^*,a}^\est(t)$, where $\kappa(c)= \bbP(\chi_{p^\rr+2}^2 \leq c)/\bbP(\chi_{p^\rr}^2 \leq c) < 1$ as defined in \citet{Li2018PNAS} under randomization with uncensored outcomes. Thus, the pointwise variance reduction at each follow-up time $t\in[0,\tau]$ is $\{1-\kappa(c)\}\bfSigma_{T,\bfB^*,a}^\est(t)^\top \bfSigma_{\bfB^*}^{-1} \bfSigma_{T,\bfB^*,a}^\est(t)$, with the relative variance reduction being 
\begin{align*}
	\frac{\{1-\kappa(c)\}\bfSigma_{T,\bfB^*,a}^\est(t)^\top \bfSigma_{\bfB^*}^{-1} \bfSigma_{T,\bfB^*,a}^\est(t)}{\Sigma_{T,a}^\est(t,t)} = \{1 - \kappa(c)\} \rho_a^\est(t).
\end{align*} 
The pointwise variance reduction factor $\rho_a^\est(t)$, defined in \eqref{eq:KM-variance-reduction}, is non-negative, because the covariance in \eqref{eq:KM-covariance} is non-zero unless $\bfZ^\rr$ is uncorrelated with the survival outcome counting process and at-risk process. Hence, compared to simple randomization, rerandomization leads to no asymptotic variance inflation and, oftentimes, improved precision for both the KM and IPCW-KM estimators. 

\subsection{Geometric illustration of the variance reduction property}

We provide a geometric illustration of the orthogonal decomposition of the limiting processes of the KM estimator under rerandomization, as established in Theorem \ref{thm:KM-IPCW-weak-convergence}. By \eqref{eq:KM-limiting-process}, the limiting process of the KM estimator admits the decomposition $\{V_a^\km(\cdot)\}^{1/2}[\{\rho_a^\km(\cdot)\}^{1/2} \cdot r_a^\km(\cdot) + \{1-\rho_a^\km(\cdot)\}^{1/2} \cdot \zeta_a^\km(\cdot)]$ uniformly over $t \in [0, \tau]$. To visualize the geometry of this process, we empirically derive the true covariance operators from a simulated super-population of $n=10^5$ units. Specifically, we generate $p^\rr = 2$ rerandomization covariates $\bfZ_i^\rr = (Z_{i1}, Z_{i2})^\top \sim \calN(\bm 0, \mathbf I_2)$. The potential event time is generated as $T_i(a) \sim \mathrm{Weibull}(\nu, \lambda_i(a))$ with shape $\nu = 1.5$ and scale $\lambda_i(a) = \exp(-\eta_i / 1.5)$, where the linear predictor is $\eta_i = 1.5 Z_{i1} + 1.5 Z_{i2}$. The censoring time $C_i\sim\calU(0, 15)$ is drawn independently of the potential event times.

\begin{figure}[htbp]
    \centering
    \includegraphics[width=\textwidth]{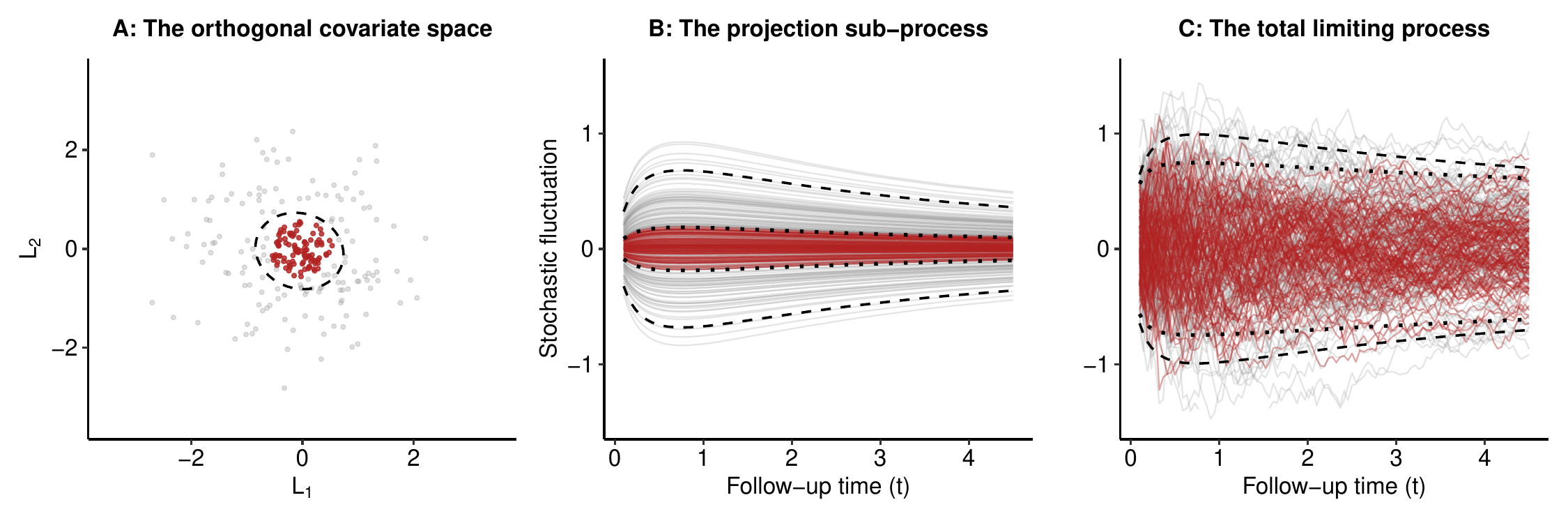}
    \caption{A geometric illustration of the effect of rerandomization on the limiting processes of the KM estimator via simulation. A total of $n=250$ independent trajectories are generated.}
    \label{fig:geometric}
\end{figure}

We set the longest follow-up time to be $\tau=5$, and the processes are evaluated over a discrete grid of 100 equally spaced points bounded within $\calT = [0.1, 4.5] \subset [0, \tau]$. We extract $V_a^\km(t)$, and $\rho_a^\km(t)$ by computing the empirical influence functions $\varphi_{T,a}^\km(t, \calO_i)$ for all units across the time grid. For the projection sub-process $r_a^\km(t) = \bfU_a^\km(t)^\top \bfL$, we simulate the standardized random vector $\bfL \sim \calN(\bm 0, \mathbf I_2)$. The balance threshold $c$ is set to the 15\% quantile of a $\chi_2^2$ distribution, yielding the constraint $\bfL^\top \bfL \leq c$. The unrestricted, orthogonal residual process $\zeta_a^\km(t)$ is generated as a mean-zero Gaussian process parameterized by the residual covariance matrix extracted from the empirical influence functions. We simulate 250 independent trajectories, partitioning them into paths that satisfy the rerandomization constraint $\bfL^\top \bfL \leq c$ and those representing the simple randomization design. The results are given in Figure \ref{fig:geometric}.

Panel A of Figure \ref{fig:geometric} shows the distribution of the standardized projection vector $\bfL \sim \calN(\bm 0, \mathbf I_2)$. The gray cloud represents the unconstrained draws under simple randomization, where extreme points correspond to trials with heavy baseline covariate imbalance on $\bfZ^\rr$. The red cluster represents the distribution under rerandomization, amputated by the condition $\bfL^\top \bfL \leq c$ (the dashed boundary). Panel B isolates the deterministic scaling of the projection sub-process, $\{V_a^\km(\cdot)\}^{1/2} \{\rho_a^\km(\cdot)\}^{1/2} \cdot r_a^\km(\cdot)$. The stochastic fluctuation of these paths is dictated entirely by the spatial spread of $\bfL$ in Panel A, modulated by the empirically derived variance envelope $V_a^\km(t)$. The dashed black lines represent the theoretical 95\% pointwise confidence intervals (CIs) under simple randomization, while the dotted black lines represent the deflated CIs under rerandomization. This panel illustrates how restricting the geometric space of covariates symmetrically truncates the structural bounds of the functional process. Panel C displays the total limiting survival process, formed by adding the unrestricted orthogonal residual $\{V_a^\km(\cdot)\}^{1/2} \{1-\rho_a^\km(\cdot)\}^{1/2} \cdot \zeta_a^\km(\cdot)$ to the projection paths from Panel B. The final 95\% uniform confidence bands (CBs) under rerandomization (dotted lines) are nested within those under simple randomization (dashed lines). This shows that rerandomization systematically reduces the stochastic variation across the entire follow-up period $t \in [0, \tau]$.

\subsection{DML estimator}

We next introduce additional regularity conditions for establishing the uniform consistency of $\widehat S_{T,a}^\dml(t)$ for $t\in[0,\tau]$ under rerandomization.
\begin{itemize}
	\item[(C3)] There exist $S_{T,a}^*(t|\bfz)$ and $S_{C,a}^*(t|\bfz)$, under simple randomization, for $\bbP$-almost all $\bfz\in\frakZ$, $\sup_{1\leq k\leq K}\sup_{t\in[0,\tau]}|\widehat S_{T,a,k}(t|\bfz)- S_{T,a}^*(t|\bfz)| \overset{p}{\to} 0$ and $\sup_{1\leq k\leq K}\sup_{t\in[0,\tau]}|\widehat S_{C,a,k}(t|\bfz)- S_{C,a}^*(t|\bfz)| \overset{p}{\to} 0$; there exists $\epsilon>0$ such that  $\inf_{t\in[0,\tau]}S_{T,a}^*(t|\bfz)\geq \epsilon$ and $\inf_{t\in[0,\tau]}S_{C,a}^*(t|\bfz) \allowbreak\geq \epsilon$, and with probability approaching 1 (over $\calI_k^c$), $\inf_{1\leq k\leq K}\inf_{t\in[0,\tau]}\allowbreak\widehat S_{T,a,k}(t|\bfz)\geq \epsilon/2$ and $\inf_{1\leq k\leq K}\inf_{t\in[0,\tau]}\widehat S_{C,a,k}(t|\bfz)\geq \epsilon/2$.
	\item[(C4)] For $\bbP$-almost all $\bfz\in\frakZ$, there exist measurable sets $\frakS_T(t|\bfz), \frakS_C(t|\bfz) \subseteq [0, t]$ such that $\frakS_T(t|\bfz)\cup\frakS_C(t|\bfz) = [0, t]$ and $S_{T,a}(u | \bfz) = S_{T,a}^*(u|\bfz)$ for all $u \in \frakS_T(t|\bfz)$ and $S_{C,a}(u|\bfz)=S_{C,a}^*(u|\bfz)$ for all $u \in \frakS_C(t|\bfz)$. 
\end{itemize}

Conditions (C3) and (C4) are adapted from Conditions (B1), (B2), and (B4) in \citet{Westling2024} for the uniform consistency of $\widehat S_{T,a}^\dml(t)$ under simple randomization or conditionally independent treatment assignment. (Detailed discussions are provided in Section S4.1 of the Supplementary Materials.) Condition (C3) requires that the estimators $\widehat S_{T,a,k}(t|\bfz)$ and $\widehat S_{C,a,k}(t|\bfz)$ are uniformly consistent for fixed limits $S_{T,a}^*(t|\bfz)$ and $S_{C,a}^*(t|\bfz)$, respectively. Similar to Condition (C2), the estimators and their limits are assumed to be bounded away from zero on $[0,\tau]$. Condition (C4) is the double robustness condition, where, essentially, either $S_{T,a}^*(t | \bfz) = S_{T,a}(t|\bfz)$ or $S_{C,a}^*(t | \bfz) = S_{C,a}(t|\bfz)$ are required on $[0,\tau]$, which is akin to the sequential doubly-robustness or $2^K$-robustness in longitudinal studies \citep[e.g.,][]{Molina2017}. Condition (C4) is a relaxation of Condition (B3) in \citet{Westling2024}, which requires either $S_{T,a}^*(t | \bfz) = S_{T,a}(t|\bfz)$ or $S_{C,a}^*(t | \bfz) = S_{C,a}(t|\bfz)$ and $\pi_a^*(\bfz) = \pi_a(\bfz)$, where $\pi_a(\bfz)=\bbP(A_i=a|\bfZ=\bfz)$. Here, we do not require the condition of $\pi_a^*(\bfz) = \pi_a(\bfz)$ because $\pi_a$ is known in a randomized experiment. The following Theorem \ref{thm:DML-consistency} gives the uniform consistency of $\widehat S_{T,a}^\dml(t)$ under rerandomization. 

\begin{theorem} \label{thm:DML-consistency}
	Under the rerandomization design, Assumptions \ref{asp:super-population}-\ref{asp:rerandomization}, and Conditions (C1), (C3), and (C4) $\sup_{t\in[0,\tau]}|\widehat S_{T,a}^\dml(t)-S_{T,a}(t)|\overset{p}{\to}0$.
\end{theorem}

We then introduce Condition (A2) for obtaining a uniform Bahadur representation and the weak convergence of $\widehat S_{T,a}^\dml(t)$ for $t\in[0,\tau]$ under rerandomization, which is an adaptation of Condition (B6) of \citet{Westling2024} under simple randomization. (Detailed discussions are provided in Section S4.1 of the Supplementary Materials.) Condition (A2) requires that both nuisance estimators converge to their corresponding true nuisance functions uniformly at a rate of $o_{\bbP}(n^{-1/4})$, which is a typical requirement in the DML literature \citep{Chernozhukov2018}.

\begin{itemize}
	\item[(A2)] Under simple randomization, $\sup_{1\leq k\leq K}\bbE\{\sup_{t\in[0,\tau]}|\widehat S_{T,a,k}(t|\bfZ)- S_{T,a}(t|\bfZ)|\} = o_{\bbP}(n^{-1/4})$ and $\sup_{1\leq k\leq K}\bbE\{\sup_{t\in[0,\tau]}|\widehat S_{C,a,k}(t|\bfZ)- S_{C,a}(t|\bfZ)|\} = o_{\bbP}(n^{-1/4})$.
\end{itemize}

\begin{theorem} \label{thm:DML-weak-convergence}
	Under the rerandomization design, Assumptions \ref{asp:super-population}-\ref{asp:rerandomization}, and Conditions (C1), (C3), (C4), and (A2), for each $t\in[0,\tau]$,
	\begin{align} \label{eq:DML-Bahadur}
		\sqrt n\left\{\widehat S_{T,a}^\dml(t)-S_{T,a}(t)\right\} = \frac{1}{\sqrt n}\sum_{i=1}^n\varphi_{T,a}^\eff(t,\calO_i) + o_{\bbP^*}(1).
	\end{align}
	The process $\{\sqrt n\{\widehat S_{T,a}^\dml(t)-S_{T,a}(t)\}:t\in[0,\tau]\}$ converges weakly to the Gaussian process $\bbG_{T,a}^\dml(\cdot)$ on $[0,\tau]$ with covariance function $\Sigma_{T,a}^\dml(s,t)=\bbE\{\varphi_{T,a}^\eff(s,\calO)\varphi_{T,a}^\eff(t,\calO)\}$.
\end{theorem}

Theorem \ref{thm:DML-weak-convergence} states that the DML estimator $\widehat S_{T,a}^\dml(t)$ admits the same Bahadur representation in \eqref{eq:DML-Bahadur} uniformly over $t\in[0,\tau]$ under rerandomization, which is consistent with the results for the KM and IPCW-KM estimators given in Theorem \ref{thm:KM-IPCW-weak-convergence}. However, rerandomization does not yield the same asymptotic variance reduction for the DML estimator, since its limiting covariance function remains unchanged. This is consistent with the discussions following Theorem \ref{thm:KM-IPCW-weak-convergence}, where the variance reduction from rerandomization hinges upon the covariance term
\begin{align} \label{eq:DML-covariance}
	\bfSigma_{T,\bfB^*,a}^\dml(t) = \bbE\left(\left[\varphi_{T,a}^\eff\{t, \calO_i(a)\}-\varphi_{T,a}^\eff\{t, \calO_i(1-a)\}\right]\left\{\bfZ_i^\rr - \bbE(\bfZ_i^\rr) \right\}\right).
\end{align}
The Neyman orthogonality of the EIF implies that $\bbE\{\varphi_{T,a}^\eff(t, \calO_i)h(\bfZ_i)\}=0$ for any integrable function $h(\bfz)$. Thus, $\bfSigma_{T,\bfB^*,a}^\dml(t)$ in \eqref{eq:DML-covariance} is equal to zero for all $t\in[0,\tau]$, since $\bfZ^\rr\subset\bfZ$. Therefore, rerandomization does not affect the asymptotic variance process of the DML estimator, as long as the estimated survival outcome function includes the rerandomization covariate. However, as we demonstrate in numerical studies, in finite samples, certain variance reduction may be observed due to improved covariate balance under rerandomization.

\section{Extensions to stratified rerandomization} \label{sec:stratified}

Results in Section \ref{sec:asymptotics} under rerandomization can be extended to the stratified rerandomization design. Specifically, let $D\in\bfZ$ be a categorical (or discretized continuous) baseline covariate encoding the randomization strata with domain $\calD=\{1,\ldots,\overline\calD\}$. We assume the number of strata $\overline\calD$ is fixed with $\bbP(D = d) > 0$ for $d \in \calD$. Assumption \ref{asp:super-population} implies that $D_1,\ldots,D_n$ are i.i.d. Following the design in \citet{Wang2023stratify}, within stratum $d$, a unit has probability $\pi_{a[d]}\in(0,1)$ to be randomized to treatment $a$, with $\pi_{1[d]}+\pi_{0[d]}=1$ for $d\in\calD$. 

The stratified rerandomization design combines stratified randomization and rerandomization as follows. First, generate $(A_1^\star,\ldots,A_n^\star)$ by stratified randomization based on observed strata $(D_1, \ldots, D_n)$ and parameter $\pi_{a[D_i]} \in (0, 1)$. Second, compute
\begin{align*}
	&\bfI_n^\star = \sum_{d\in\calD}\frac{n_{[d]}}{n}\sum_{i=1}^n\bbone(D_i=d)\left\{ \frac{A_i^\star\bfZ_i^\rr}{n_{1[d]}^\star} - \frac{(1-A_i^\star)\bfZ_i^\rr}{n_{0[d]}^\star}\right\},\displaybreak[0]\\
	&\widehat{\Var}(\bfI_n^\star) = \sum_{d\in\calD}\frac{n_{[d]}}{n}\cdot\frac{1}{n_{1[d]}^\star n_{0[d]}^\star}\sum_{i=1}^n\bbone(D_i=d)\left(\bfZ_i^\rr-\overline\bfZ_{[d]}^\rr\right)\left(\bfZ_i^\rr-\overline\bfZ_{[d]}^\rr\right)^\top,
\end{align*}
where $n_{[d]}=\sum_{i=1}^n\bbone(D_i=d)$, $n_{a[d]}^\star=\sum_{i=1}^n\bbone(D_i=d)\bbone(A_i^\star=a)$ and $\overline\bfZ_{[d]}^\rr=n_{[d]}^{-1}\sum_{i=1}^n\bbone(D_i=d)\bfZ_i^\rr$. Lastly, if $\bfI_n^{\star\top}\{\widehat{\Var}(\bfI_n^\star)\}^{-1}\bfI_n^\star < c$ for a pre-specified balance threshold $c$, set the final treatment $(A_1,\ldots,A_n)$ to $(A_1^\star,\ldots,A_n^\star)$; otherwise, repeat the steps until obtaining the first randomization scheme that satisfies the balance criterion. Without loss of generality, we assume $D\notin\bfZ^\rr$, since alternatively, rerandomization is effectively controlling for imbalance on $\bfZ^\rr \setminus D$. Similar to $\bfB_n^*$ in \eqref{eq:B-star} without the stratification, define
\begin{align*}
	&\bfB_{[d],n_{[d]}}^\star = \frac{1}{\sqrt{n_{[d]}}} \sum_{i:D_i=d} \frac{1}{\pi_{1[D_i]}\pi_{0[D_i]}} \left(A_i^\star-\pi_{1[D_i]}\right)\left(\bfZ_i^\rr - \overline\bfZ_{[D_i]}^\rr\right), \displaybreak[0]\\
	&\bfSigma_{\bfZ^\rr[d],n_{[d]}}=\frac{1}{n_{[d]}}\sum_{i:D_i=d}\left(\bfZ_i^\rr-\overline\bfZ_{[d]}^\rr\right)\left(\bfZ_i^\rr-\overline\bfZ_{[d]}^\rr\right)^\top.
\end{align*}
Then, let $\bfB_n^\star=\sum_{d\in\calD}\gamma_{[d]}^{1/2}\bfB_{[d],n_{[d]}}^\star$, where $\gamma_{[d]}=\bbP(D=d)$, and we have $\sqrt n\bfI_n^\star = \bfB_n^\star + o_{\bbP}(1)$. By Lemma S3 in Section S5.1 of the Supplementary Materials, we have 
\begin{align*}
	\calE_n^s\equiv\{\bfI_n^{\star\top} \widehat\Var(\bfI_n^\star)^{-1} \bfI_n^\star < c\} = \{\bfB_n^{\star\top}\bfSigma_{\bfB^\star,n}^{-1}\bfB_n^\star\} + o_{\bbP}(1),
\end{align*}
where $\bfSigma_{\bfB^\star,n}=\sum_{d\in\calD}(\pi_{1[d]}\pi_{0[d]})^{-1}\bfSigma_{\bfZ^\rr[d],n_{[d]}}$. Thus, asymptotically, conditioning on $\calE_n^s$ is equivalent to conditioning on $\bfB_n^\star$ lying in the ellipsoid $\{ \bfx \in \bbR^{p^\rr} : \bfx^\top \bfSigma_{\bfB^\star,n}^{-1} \bfx < c \}$, with $\calE^s\equiv\lim_{n\to\infty}\calE_n^s=\{ \bfx \in \bbR^{p^\rr} : \bfx^\top \bfSigma_{\bfB^\star}^{-1} \bfx < c \}$, where $\bfSigma_{\bfB^\star}=\sum_{d\in\calD}(\pi_{1[d]}\pi_{0[d]})^{-1}\bfSigma_{\bfZ^\rr[d]}$ and $\bfSigma_{\bfZ^\rr[d]}\equiv\Cov(\bfZ_i^\rr|D_i=d)$. Theoretical results of the KM and IPCW-KM estimators under stratified rerandomization are given in Theorems \ref{thm:KM-IPCW-consistency-stratified} and \ref{thm:KM-IPCW-weak-convergence-stratified}.

\begin{theorem} \label{thm:KM-IPCW-consistency-stratified}
	(i) Under the stratified rerandomization design, Assumptions \ref{asp:super-population}, \ref{asp:censor}(i) and (iii), and \ref{asp:rerandomization}, and Condition (C1), $\sup_{t\in[0,\tau]}|\widehat S_{T,a}^\km(t)-S_{T,a}(t)|\overset{p}{\to}0$. (ii) Under the stratified rerandomization design, Assumptions \ref{asp:super-population}, \ref{asp:censor}(ii) and (iii), and Conditions (C1) and (C2), $\sup_{t\in[0,\tau]}|\widehat S_{T,a}^\ipcw(t)-S_{T,a}(t)|\overset{p}{\to}0$.
\end{theorem}

\begin{theorem} \label{thm:KM-IPCW-weak-convergence-stratified}
	Under the stratified rerandomization design, Assumptions \ref{asp:super-population}-\ref{asp:rerandomization}, and Conditions (C1) and (A1), for $\est=\km,\ipcw$ and all $t\in[0,\tau]$,
	\begin{align} \label{eq:KM-Bahadur-stratified}
		\sqrt n\left\{\widehat S_{T,a}^\est(t)-S_{T,a}(t)\right\} = \frac{1}{\sqrt n}\sum_{i=1}^n\varphi_{T,a}^\est(t,\calO_i) + o_{\bbP^*}(1).
	\end{align}
	The process $\{\sqrt n\{\widehat S_{T,a}^\est(t)-S_{T,a}(t)\}:t\in[0,\tau]\}$ converges weakly to a mean-zero process 
	\begin{align} \label{eq:KM-limiting-process-stratified}
		\left\{V_a^\est(\cdot)\right\}^{1/2}  \left[\left\{\rho_a^\est(\cdot)\right\}^{1/2} \cdot r_a^\est(\cdot) + \left\{1 - \rho_a^\est(\cdot)\right\}^{1/2} \cdot \zeta_a^\est(\cdot)\right]
	\end{align}
	on $[0,\tau]$. Here, $r_a^\est(\cdot)$ and $\zeta_a^\est(\cdot)$ are two mean-zero, tight processes with $r_a^\est(\cdot)\perp \zeta_a^\est(\cdot)$, $V_a^\est(t)=\Sigma_{T,a}^\est(t,t)$, and 
	\begin{align*} 
		\rho_a^\est(t) = \frac{\bfSigma_{T,\bfB^\star,a}^\est(t)^\top \bfSigma_{\bfB^\star}^{-1} \bfSigma_{T,\bfB^\star,a}^\est(t)}{\Sigma_{T,a}^\est(t,t)},
	\end{align*}
	where 
	\begin{align} \label{eq:KM-covariance-stratified}
		\bfSigma_{T,\bfB^\star,a}^\est(t) = \bbE\left\{\bbE\left(\left[\varphi_{T,a}^\est\left\{t, \calO_i^{(D_i)}(a)\right\}-\varphi_{T,a}^\est\left\{t, \calO_i^{(D_i)}(1-a)\right\}\right]\left\{\bfZ_i^\rr - \bbE(\bfZ_i^\rr|D_i) \right\}\middle|D_i\right)\right\},
	\end{align}    
	where $\calO_i^{(D_i)}(a)$ signifies the observed stratum $D_i$ in $\calO_i(a)$. The process $r_a^\est(\cdot)=\bfU_a^\est(\cdot)^\top\bfL$, where $\bfU_a^\est(\cdot)=\bfalpha_a^\est(\cdot)/\|\bfalpha_a^\est(\cdot)\|$ with $\bfalpha_a^\est(\cdot)=\bfSigma_{\bfB^\star}^{-1/2} \bfSigma_{T,\bfB^\star,a}^\est(\cdot)$; for $t\in[0,\tau]$, $\Var\{r_a^\est(t)\} = \kappa(c) = \bbE(L_1^2 | \bfL^\top \bfL < c)$, with $L_1$ being the first element of $\bfL\sim\calN(\bm 0,\mathbf I_{p^\rr})$ conditioned on $\bfL^\top \bfL \leq c$, and for $s, t \in [0,\tau]$, $\Cov\{r_a^\est(s), r_a^\est(t)\}  = \kappa(c) \bfU_a^\est(s)^\top \bfU_a^\est(t)$. The process $\zeta_a^\est(\cdot)$ is a standardized Gaussian process on $[0,\tau]$, with 
	\begin{align*}
		\Cov\left\{\zeta_a^\est(s),\zeta_a^\est(t)\right\} = \frac{\Sigma_{T,a}^\est(s,t)-\bfSigma_{T,\bfB^\star,a}^\est(s)^\top \bfSigma_{\bfB^\star}^{-1} \bfSigma_{T,\bfB^\star,a}^\est(t)}{[V_a^\est(s)\{1 - \rho_a^\est(s)\}]^{1/2}[V_a^\est(t)\{1 - \rho_a^\est(t)\}]^{1/2}},
	\end{align*}
	for $s, t \in [0,\tau]$. 
\end{theorem}

Theorems \ref{thm:KM-IPCW-consistency-stratified} and \ref{thm:KM-IPCW-weak-convergence-stratified} are analogous to Theorems \ref{thm:KM-IPCW-consistency} and \ref{thm:KM-IPCW-weak-convergence} concerning the uniform consistency, Bahadur representation, and weak convergence of the KM and IPCW-KM estimators under stratified rerandomization. Stratified rerandomization induces a variance reduction in $\widehat S_{T,a}^\km(t)$ and $\widehat S_{T,a}^\ipcw(t)$ that is comparable to that achieved under standard rerandomization. The principal distinction arises in \eqref{eq:KM-covariance-stratified}, where, under stratified rerandomization, the covariance structure is further adjusted to account for the stratification variable $D$. For completeness, we next present the asymptotic results for the DML estimator under stratified rerandomization.

\begin{theorem} \label{thm:DML-consistency-stratified}
	Under the rerandomization design, Assumptions \ref{asp:super-population}-\ref{asp:rerandomization}, and Conditions (C1), (C3), and (C4) $\sup_{t\in[0,\tau]}|\widehat S_{T,a}^\dml(t)-S_{T,a}(t)|\overset{p}{\to}0$.
\end{theorem}

\begin{theorem} \label{thm:DML-weak-convergence-stratified}
	Under the rerandomization design, Assumptions \ref{asp:super-population}-\ref{asp:rerandomization}, and Conditions (C1), (C3), (C4), and (A2), for each $t\in[0,\tau]$,
	\begin{align} \label{eq:DML-Bahadur-stratified}
		\sqrt n\left\{\widehat S_{T,a}^\dml(t)-S_{T,a}(t)\right\} = \frac{1}{\sqrt n}\sum_{i=1}^n\varphi_{T,a}^\eff(t,\calO_i) + o_{\bbP^*}(1).
	\end{align}
	The process $\{\sqrt n\{\widehat S_{T,a}^\dml(t)-S_{T,a}(t)\}:t\in[0,\tau]\}$ converges weakly to the Gaussian process $\bbG_{T,a}^\dml(\cdot)$ on $[0,\tau]$ with covariance function $\Sigma_{T,a}^\dml(s,t)=\bbE\{\varphi_{T,a}^\eff(s,\calO)\varphi_{T,a}^\eff(t,\calO)\}$.
\end{theorem}

Theorems \ref{thm:DML-consistency-stratified} and \ref{thm:DML-weak-convergence-stratified} establish the uniform consistency, Bahadur representation, and weak convergence of the DML estimator under stratified rerandomization. Analogous to Theorems \ref{thm:DML-consistency} and \ref{thm:DML-weak-convergence}, stratified rerandomization does not provide additional variance reduction to the DML estimator because of the Neyman orthogonality of the EIF, as long as the stratified variable is contained in the survival outcome regression covariate adjustment set, or equivalently, $D\in\bfZ$.

\section{Numerical studies} \label{sec:simulation}

\subsection{Simulation design}

We conduct simulation studies to illustrate our asymptotic results, considering three randomization procedures: simple randomization, rerandomization, and stratified rerandomization. We generate the covariate vector $\bfZ_i = (Z_{i1}, Z_{i2}, D_i)^\top$, where $Z_{i1} \sim \calN(1, 1)$, $D_i | Z_{i1} \sim \calB(p_i)$, a Bernoulli random variable with $p_i \equiv \bbP(D_i = 1 | Z_{i1}) = 0.4 + 0.2 \bbone(X_{i1} < 1)$, and $Z_{i2} | Z_{i1}, D_i \sim \calN(0, 1)$. The rerandomization covariates $\bfZ_i^\rr = (Z_{i1}, Z_{i2})^\top$. The treatment assignment $A_i \in \{0, 1\}$ is generated with $\pi_1=\pi_0 = 0.5$ under three designs: (i) the simple randomization: draw $A_i \sim \calB(0.5)$ independently; (ii) rerandomization: draw $A_i^* \sim \calB(0.5)$, compute the Mahalanobis distance $\bfI_n^{\ast\top}\{\widehat{\Var}(\bfI_n^*)\}^{-1}\bfI_n^*$ based on $\bfZ^\rr$, and accept the allocation if $\bfI_n^{\ast\top}\{\widehat{\Var}(\bfI_n^*)\}^{-1}\bfI_n^* < 1.83$ (the approximate 40th percentile of a $\chi^2_2$ distribution, yielding a 40\% acceptance rate); and (iii) stratified rerandomization: draw $A_i^\star \sim \calB(0.5)$ within the strata defined by $D_i$, compute the stratified Mahalanobis distance $\bfI_n^{\star\top}\{\widehat{\Var}(\bfI_n^\star)\}^{-1}\bfI_n^\star$ on $\bfZ^\rr$, and accept if $\bfI_n^{\star\top}\{\widehat{\Var}(\bfI_n^\star)\}^{-1}\bfI_n^\star < 1.83$.

For outcome and censoring time generation, we consider two scenarios: in scenario I (proportional hazards with covariate-dependent censoring), the outcome time $T_i(a) \sim \mathrm{Weibull}\{1.5, \lambda_i(a)^{-1/1.5}\}$, where $\lambda_i(a) = 0.1 \exp(-0.5 a + 0.5 Z_{i1} - 0.5 Z_{i2} + 0.5 D_i + 0.5 a Z_{i1} - 0.5 a Z_{i2})$ for $a=0,1$, and the censoring time $C_i \sim \text{Exp}\{0.05 \exp(0.2 Z_{i1} + 0.2 D_i)\}$; in scenario II (accelerated failure time with completely independent censoring), the outcome time $\log T_i(a) \sim \calN(\mu_i(a),\sigma^2)$, where $\mu_i(a) = -1 - 0.5a + Z_{i1} - Z_{i2} + D_i$ for $a=0,1$, and the censoring time $C_i \sim \calU(0, 20)$. In both scenarios, we set $\tau = 5$. The parameters described above are calibrated such that the resulting mean observed event rate is approximately 50\% in Scenario I and 70\% in Scenario II, while the proportion of censoring prior to $\tau = 5$ lies between 20\% and 30\% in both scenarios. The parameters of interest are $S_{T,1}(t) = \bbP\{T(1)>t^*\}$ at $t^*=1,2,3,4$, where the true values are approximated via Monte Carlo integration over a simulated super-population of $n = 10^6$ units using the true data-generating process parameters. In Scenario I, we consider the IPCW-KM and DML estimators due to covariate-dependent censoring. In Scenario II, we consider the KM and DML estimators, as the censoring is completely independent. For the IPCW-KM estimator, we fit a Cox PH model for the censoring time, and for the DML estimator, we adopt a five-fold cross-fitting scheme given in Section \ref{sec:DML} by setting $K=5$. The implementation of the DML estimator is conducted via the {\tt R} package {\tt survSuperLearner}. We execute this for sample sizes $n = 100$ and $400$ with 1000 replicates.

\subsection{Pointwise and uniform inference}

The pointwise and uniform inference procedures follow \citet{Westling2024}, and a detailed methodological discussion of variance estimation is not particularly interesting, given its similarity to prior work. We thus provide descriptions of the procedures for obtaining pointwise CIs and the uniform CBs. 

To obtain pointwise CIs, for $\est=\km,\ipcw$, define $\sigma_{T,a}^{\est,2}(s,t)\equiv\Sigma_{T,a}^\est(s,t)$ under simple randomization, $\sigma_{T,a}^{\est,2}(s,t)\equiv\Sigma_{T,a}^\est(s,t)-\{1-\kappa(c)\}\bfSigma_{T,\bfB^*,a}^\est(s)^\top \bfSigma_{\bfB^*}^{-1} \bfSigma_{T,\bfB^*,a}^\est(t)$ under rerandomization, and $\sigma_{T,a}^{\est,2}(s,t)\equiv\Sigma_{T,a}^\est(s,t)-\{1-\kappa(c)\}\bfSigma_{T,\bfB^\star,a}^\est(s)^\top \bfSigma_{\bfB^\star}^{-1} \bfSigma_{T,\bfB^\star,a}^\est(t)$ under stratified rerandomization. Then, a natural estimator of $\sigma_{T,a}^{\est,2}(s,t)$ is $\widehat\sigma_{T,a}^{\est,2}(s,t)\equiv\widehat\Sigma_{T,a}^\est(s,t)$ for the simple randomization design, $\widehat\sigma_{T,a}^{\est,2}(s,t)\equiv\widehat\Sigma_{T,a}^\est(s,t)-\{1-\kappa(c)\}\widehat\bfSigma_{T,\bfB^*,a}^\est(s)^\top \widehat\bfSigma_{\bfB^*}^{-1} \widehat\bfSigma_{T,\bfB^*,a}^\est(t)$ for the rerandomization design, and $\widehat\sigma_{T,a}^{\est,2}(s,t)\equiv\widehat\Sigma_{T,a}^\est(s,t)-\{1-\kappa(c)\}\widehat\bfSigma_{T,\bfB^\star,a}^\est(s)^\top \widehat\bfSigma_{\bfB^\star}^{-1} \widehat\bfSigma_{T,\bfB^\star,a}^\est(t)$ for the stratified rerandomization design. For the DML estimator, define $\sigma_{T,a}^{\dml,2}(s,t)\equiv\Sigma_{T,a}^\dml(s,t)$ with $\widehat\sigma_{T,a}^{\dml,2}(s,t)\equiv\widehat\Sigma_{T,a}^\dml(s,t)$, since neither rerandomization nor stratified rerandomization alters its covariance function. Note that $\widehat\sigma_{T,a}^{\dml,2}(s,t)$ is obtained following the same cross-fitting scheme given in Section \ref{sec:DML} with $K=5$. The specific forms of $\widehat\sigma_{T,a}^{\est,2}(s,t)$ ($\est=\km,\ipcw$) under different designs are obtained by replacing the population expectations in the expressions with their empirical counterparts. For $\widehat\sigma_{T,a}^{\dml,2}(s,t)$, we recycle the point estimates with the previous cross-fitted EIFs, i.e., $\widehat\sigma_{T,a}^{\dml,2}(s,t)=\bbP_{n,K}\{\widehat\varphi_{T,a}^\eff(s,\calO)\widehat\varphi_{T,a}^\eff(t,\calO)\}$, where $\widehat\varphi_{T,a}^\eff(t,\calO)=\widehat\varphi_{T,a}^{\eff,\ast}(t,\calO) + \widehat S_{T,a}^\dml(t)$.

With $\widehat\sigma_{T,a}^{\est,2}(t)\equiv\widehat\sigma_{T,a}^{\est,2}(t,t)$, the pointwise Wald-type asymptotic $(1 - \alpha)$-level CI for $S_{T,a}(t)$ is $\widehat S_{T,a}^\est(t) \pm z_{1-\alpha/2}\cdot\widehat\sigma_{T,a}^\est(t)/\sqrt n$, where $z_{1-\alpha/2}$ is the $(1-\alpha/2)$-quantile of the standard normal distribution. As suggested in \citet{Westling2024}, constructing Wald-type intervals on the logistic probability scale can improve finite-sample coverage. Specifically, define $\mathrm{expit}(x) \equiv \exp(x)/\{1 + \exp(x)\}$ for $x \in \bbR$ and $\mathrm{logit}(u) \equiv \log(u) - \log(1 - u)$ for $u \in (0, 1)$, with $\widetilde\sigma_{T,a}^\est(t) \equiv \widehat\sigma_{T,a}^\est(t)/[\widehat S_{T,a}^\est(t)\{1-\widehat S_{T,a}^\est(t)\}]$. Then, the transformed Wald-type interval $[l_{T,a}^\est(t), u_{T,a}^\est(t)] \equiv \mathrm{expit}[\mathrm{logit}\{\widehat S_{T,a}^\est(t)\} \pm z_{1-\alpha/2}\cdot \widetilde\sigma_{T,a}^\est(t)/\sqrt n]$. If $\widehat S_{T,a}^\est(t)=0$, set $l_{T,a}^\est(t)=0$ and $u_{T,a}^\est(t) = \min_s\{u_{T,a}^\est(s) : u_{T,a}^\est(s) > 0\}$, whereas if $\widehat S_{T,a}^\est(t)=1$, set $l_{T,a}^\est(t) = \min_s\{l_{T,a}^\est(s) : l_{T,a}^\est(s) > 0\}$ and $u_{T,a}^\est(t)=1$.

For uniform inference, we construct fixed-width CBs as $\widehat S_{T,a}^\est(t) \pm \widehat c_{a,\alpha}^\est/\sqrt{n}$, where $\widehat c_{a,\alpha}^\est$ is any consistent estimator of the $(1 - \alpha)$-quantile of the supremum of the absolute value of the limiting mean-zero Gaussian process to which $\{\sqrt n\{\widehat S_{T,a}^\est(t)-S_{T,a}(t)\} : t \in \calT\}$ weakly converges, for a restricted compact and continuous time domain $\calT \subset [0, \tau]$. To obtain $\widehat c_{a,\alpha}^\est$, we simulate sample paths of a mean-zero Gaussian process with covariance function $(s, t) \mapsto \widehat\sigma_{T,a}^{\est,2}(s,t)$, and then set $\widehat c_{a,\alpha}^\est$ as the sample $(1-\alpha)$-quantile of the uniform norm over $\calT$ of these sample paths. Specifically, in our simulations, we evaluate simultaneous coverage over $\mathcal{T} = [1,4]$, obtaining the uniform CBs using 50 equally spaced time points over this interval. The monotonicity of the CBs is ensured via isotonic regression \citep{Westling2020}.

\subsection{Simulation results}

Simulation results are reported in Tables \ref{tab:scenario-1-n-100} and \ref{tab:scenario-2-n-100} for settings with sample size $n=100$ and in Web Tables 1 and 2 in Section S7 of the Supplementary Materials for sample size $n=400$. All estimators are asymptotically consistent, and the empirical coverage probabilities of the pointwise 95\% CIs attain their nominal levels. The uniform 95\% CBs are conservative across the considered scenarios. These findings provide empirical corroboration for Theorems \ref{thm:KM-IPCW-consistency}, \ref{thm:DML-consistency}, \ref{thm:KM-IPCW-consistency-stratified}, and \ref{thm:DML-consistency-stratified}.

In Scenario I, under covariate-dependent censoring, the IPCW estimator yields larger empirical standard errors (ESEs) under simple randomization than under rerandomization and stratified rerandomization, demonstrating that rerandomization and stratified rerandomization reduce the uncertainty of the IPCW-KM estimator. In Scenario II, under completely independent censoring, the KM estimator exhibits the same pattern, indicating that the variance-reduction properties of rerandomization and stratified rerandomization persist in this setting. These findings provide empirical confirmation of Theorems \ref{thm:KM-IPCW-weak-convergence} and \ref{thm:KM-IPCW-weak-convergence-stratified}. The variance reduction effect is more pronounced in scenario II because of the stronger effect of $\bfZ$ and $D$ in the data-generating process.

In comparison, under both scenarios, the DML estimators exhibit comparable ESEs under all three randomization designs, corroborating Theorems \ref{thm:DML-weak-convergence} and \ref{thm:DML-weak-convergence-stratified}, which state that rerandomization and stratified rerandomization do not reduce the variance of the DML estimator due to the Neyman orthogonality of the EIF. Interestingly, the DML estimator is empirically more efficient than the IPCW-KM or the KM estimator under rerandomization and stratified rerandomization, yielding smaller ESEs across the considered scenarios. This indicates that, in nonlinear regression settings, covariate adjustment at the analysis stage can remain critically important and can dominate the efficiency gains attributable to design-based adjustment via rerandomization.

\begin{table}[htbp]
	\centering
	\caption{Simulation results from scenario I with sample size $n=100$. IPCW: the IPCW-KM estimator. DML: the DML estimator. SRS: simple randomization, ReM: rerandomization, SReM: stratified rerandomization. Bias: absolute value of the bias. ESE: empirical standard error. ASE: average standard error. ECP: empirical coverage percentage of the pointwise 95\% CI. U-ECP: empirical coverage percentage of the uniform 95\% CB.} \label{tab:scenario-1-n-100}
	\begin{tabular}{ll crccc c}
		\toprule
		\multirow{2}{*}{Method} & \multirow{2}{*}{Design} & \multicolumn{5}{c}{Pointwise metrics at time $t$} & \multirow{2}{*}{U-ECP} \\
		\cmidrule{3-7}
		& & Time & Bias & ESE & ASE & ECP & \\
		\midrule
		\multirow{12}{*}{IPCW} & \multirow{4}{*}{SRS} 
		& 1 & .002 & .065 & .063 & .946 & \multirow{4}{*}{.958} \\
		& & 2 & .003 & .074 & .071 & .945 & \\
		& & 3 & .006 & .072 & .070 & .952 & \\
		& & 4 & .010 & .068 & .066 & .946 & \\
		\cmidrule{2-8}
		& \multirow{4}{*}{ReM} 
		& 1 & .004 & .061 & .060 & .957 & \multirow{4}{*}{.956} \\
		& & 2 & .004 & .067 & .067 & .949 & \\
		& & 3 & .009 & .066 & .066 & .946 & \\
		& & 4 & .012 & .064 & .063 & .939 & \\
		\cmidrule{2-8}
		& \multirow{4}{*}{SReM} 
		& 1 & .004 & .064 & .060 & .944 & \multirow{4}{*}{.954} \\
		& & 2 & .007 & .073 & .067 & .931 & \\
		& & 3 & .008 & .071 & .066 & .936 & \\
		& & 4 & .009 & .068 & .063 & .925 & \\
		\midrule
		\multirow{12}{*}{DML} & \multirow{4}{*}{SRS} 
		& 1 & .013 & .062 & .059 & .928 & \multirow{4}{*}{.959} \\
		& & 2 & .019 & .068 & .066 & .938 & \\
		& & 3 & .021 & .065 & .064 & .945 & \\
		& & 4 & .021 & .063 & .060 & .945 & \\
		\cmidrule{2-8}
		& \multirow{4}{*}{ReM} 
		& 1 & .012 & .060 & .059 & .948 & \multirow{4}{*}{.956} \\
		& & 2 & .020 & .066 & .066 & .943 & \\
		& & 3 & .021 & .064 & .064 & .944 & \\
		& & 4 & .022 & .063 & .061 & .941 & \\
		\cmidrule{2-8}
		& \multirow{4}{*}{SReM} 
		& 1 & .012 & .063 & .059 & .941 & \multirow{4}{*}{.956} \\
		& & 2 & .017 & .069 & .065 & .936 & \\
		& & 3 & .021 & .067 & .064 & .939 & \\
		& & 4 & .024 & .065 & .060 & .941 & \\
		\bottomrule
	\end{tabular}
\end{table}

\begin{table}[htbp]
	\centering
	\caption{Simulation results from scenario II with sample size $n=100$. KM: the KM estimator. DML: the DML estimator. SRS: simple randomization, ReM: rerandomization, SReM: stratified rerandomization. Bias: absolute value of the bias. ESE: empirical standard error. ASE: average standard error. ECP: empirical coverage percentage of the pointwise 95\% CI. U-ECP: empirical coverage percentage of the uniform 95\% CB.} \label{tab:scenario-2-n-100}
	\begin{tabular}{ll crccc c}
		\toprule
		\multirow{2}{*}{Method} & \multirow{2}{*}{Design} & \multicolumn{5}{c}{Pointwise metrics at time $t$} & \multirow{2}{*}{U-ECP} \\
		\cmidrule{3-7}
		& & Time & Bias & ESE & ASE & ECP & \\
		\midrule
		\multirow{12}{*}{KM} & \multirow{4}{*}{SRS} 
		& 1 & .006 & .072 & .069 & .934 & \multirow{4}{*}{.945} \\
		& & 2 & .004 & .069 & .064 & .938 & \\
		& & 3 & .002 & .064 & .058 & .934 & \\
		& & 4 & .002 & .058 & .052 & .930 & \\
		\cmidrule{2-8}
		& \multirow{4}{*}{ReM} 
		& 1 & .005 & .064 & .063 & .957 & \multirow{4}{*}{.952} \\
		& & 2 & .006 & .063 & .060 & .950 & \\
		& & 3 & .003 & .060 & .054 & .926 & \\
		& & 4 & .002 & .056 & .049 & .916 & \\
		\cmidrule{2-8}
		& \multirow{4}{*}{SReM} 
		& 1 & .007 & .066 & .063 & .943 & \multirow{4}{*}{.961} \\
		& & 2 & .006 & .063 & .060 & .942 & \\
		& & 3 & .001 & .060 & .054 & .924 & \\
		& & 4 & .002 & .056 & .049 & .936 & \\
		\midrule
		\multirow{12}{*}{DML} & \multirow{4}{*}{SRS} 
		& 1 & .016 & .058 & .058 & .947 & \multirow{4}{*}{.938} \\
		& & 2 & .022 & .056 & .055 & .939 & \\
		& & 3 & .026 & .055 & .050 & .938 & \\
		& & 4 & .028 & .051 & .046 & .959 & \\
		\cmidrule{2-8}
		& \multirow{4}{*}{ReM} 
		& 1 & .017 & .057 & .058 & .954 & \multirow{4}{*}{.949} \\
		& & 2 & .020 & .056 & .055 & .948 & \\
		& & 3 & .024 & .053 & .050 & .935 & \\
		& & 4 & .026 & .050 & .046 & .969 & \\
		\cmidrule{2-8}
		& \multirow{4}{*}{SReM} 
		& 1 & .015 & .058 & .058 & .953 & \multirow{4}{*}{.954} \\
		& & 2 & .020 & .055 & .055 & .951 & \\
		& & 3 & .025 & .053 & .050 & .940 & \\
		& & 4 & .026 & .051 & .046 & .964 & \\
		\bottomrule
	\end{tabular}
\end{table}

\subsection{An illustrative real data example}

We additionally illustrate the theoretical developments using a real data example based on data from a randomized clinical trial evaluating hormonal therapy and chemotherapy duration in patients with node-positive breast cancer. The trial was conducted by the German Breast Cancer Study Group \citep[GBCSG;][]{Schumacher1994}, and the dataset analyzed here is available as \texttt{GBSG2} in the \texttt{R} package \texttt{TH.data}. The dataset comprises 686 patients, with recurrence-free survival time serving as the primary outcome ($T_i$). The original study employed a $2\times2$ factorial design, with the two randomized factors being hormonal therapy (yes vs. no) and chemotherapy duration (short vs. long). However, the chemotherapy duration variable is not included in the \texttt{GBSG2} dataset. Consistent with prevailing practice in the literature, analyses typically concentrate on the marginal comparison of hormonal therapy, implicitly averaging over chemotherapy duration. This approach is valid under the assumption of no interaction between the two treatment factors. In line with this convention, we restrict attention to the comparison between hormonal therapy ($a=1$) and no hormonal therapy ($a=0$); in the available data, 246 patients (35.9\%) received hormonal therapy.

The dataset includes seven baseline prognostic covariates (age, menopausal status, tumor size, tumor grade, number of positive lymph nodes, progesterone receptor level, and estrogen receptor level). Among these, tumor size and tumor grade are categorical variables. A logarithmic transformation is applied to tumor size, number of positive nodes, progesterone receptor, and estrogen receptor because of the pronounced right skewness in their empirical distributions. Of the 686 patients, 299 (43.6\%) exhibit right-censored outcomes. The maximum observed event time is 2659 days, and the maximum observed censoring time is 2456 days. We assess the impact of rerandomization and stratified rerandomization by estimating the percentage reduction in the asymptotic variances of the IPCW-KM estimator (assuming conditionally independent censoring) and KM estimator (assuming completely independent censoring) for $S_{T,1}(t)$ that would have been achieved had the observed treatments been assigned under these two designs. The number of positive nodes and progesterone receptor are chosen as rerandomization covariates, and tumor grade is used as the stratification covariate. We estimate the survival probability under hormonal therapy and its pointwise variance at four time points, $t = 500, 1000, 1500,$ and $2000$ days. For completeness, we also compute the DML estimator for comparison. The results are reported in Table \ref{tab:real-data-gbsg2}.

\begin{table}[htbp]
	\centering
	\caption{Empirical estimates, standard errors, and variance reduction from the GBSG2 data illustration. Estimates are provided with empirical standard errors in parentheses. KM: the KM estimator. IPCW-KM: the IPCW-KM estimator. DML: the DML estimator. SRS: simple randomization, ReM: rerandomization, SReM: stratified rerandomization. The variance reduction is calculated relative to the unadjusted variance under simple randomization.} \label{tab:real-data-gbsg2}
	\begin{tabular}{ll cccc}
		\toprule
		\multirow{2}{*}{Estimator} & \multirow{2}{*}{Design} & \multicolumn{4}{c}{Follow-up time $t$ (days)} \\
		\cmidrule{3-6}
		& & 500 & 1000 & 1500 & 2000 \\
		\midrule
		\multicolumn{6}{c}{Point estimates (standard errors)} \\
		\midrule
		\multirow{3}{*}{KM} 
		& SRS  & .898 (.014) & .723 (.021) & .627 (.024) & .526 (.030) \\
		& ReM  & .898 (.014) & .723 (.021) & .627 (.023) & .526 (.029) \\
		& SReM & .898 (.014) & .723 (.021) & .627 (.023) & .526 (.029) \\
		\midrule
		\multirow{3}{*}{IPCW-KM} 
		& SRS  & .898 (.014) & .723 (.021) & .629 (.024) & .513 (.035) \\
		& ReM  & .898 (.014) & .723 (.021) & .629 (.023) & .513 (.034) \\
		& SReM & .898 (.014) & .723 (.021) & .629 (.023) & .513 (.034) \\
		\midrule
		\multirow{3}{*}{DML}     
		& SRS  & .898 (.014) & .723 (.021) & .622 (.023) & .523 (.030) \\
		& ReM  & .898 (.014) & .723 (.021) & .622 (.023) & .523 (.030) \\
		& SReM & .898 (.014) & .723 (.021) & .622 (.023) & .523 (.030) \\
		\midrule
		\multicolumn{6}{c}{Relative variance reduction (\%)} \\
		\midrule
		\multirow{2}{*}{KM}      
		& ReM  & 4.21 & 6.41 & 6.33 & 4.55 \\
		& SReM & 4.51 & 5.50 & 5.42 & 3.90 \\
		\midrule
		\multirow{2}{*}{IPCW-KM} 
		& ReM  & 4.20 & 6.57 & 6.55 & 3.30 \\
		& SReM & 4.52 & 5.63 & 5.61 & 2.85 \\
		\bottomrule
	\end{tabular}
\end{table}

Table \ref{tab:real-data-gbsg2} shows that the point estimates for the KM, IPCW-KM, and DML estimators remain consistent across all designs. Under rerandomization and stratified rerandomization, the KM and IPCW-KM estimators achieve relative variance reductions of 2.85\% to 6.57\% compared with simple randomization, demonstrating that rerandomization and stratified rerandomization do not inflate the asymptotic variances.

\section{Discussion} \label{sec:conclusion}

The uniform limiting processes established in Theorems \ref{thm:KM-IPCW-weak-convergence} and \ref{thm:KM-IPCW-weak-convergence-stratified} reveal a fundamental geometric connection between design-stage constraints and analysis-stage efficiency. The variance reduction achieved by rerandomization is mathematically equivalent to clamping the stochastic variation of the estimator's projection onto the restricted covariate subspace. For the KM and IPCW-KM estimators, this geometric clamping deflates the uniform CBs. However, as demonstrated in Theorems \ref{thm:DML-weak-convergence} and \ref{thm:DML-weak-convergence-stratified}, this geometric advantage is nullified when utilizing the DML estimator. Because the EIF is Neyman-orthogonal to the baseline covariates, its projection onto the rerandomization subspace is asymptotically zero. Consequently, when the analysis-stage estimator achieves the semiparametric efficiency bound, restricting the design space offers no further asymptotic variance reduction. 

In practice, however, the KM estimator remains the most conventional and widely reported choice for survival analysis in clinical trials, and this operational boundary carries an important interpretive caveat. When rerandomization or any other form of constrained randomization has been employed, interval estimates derived from the unadjusted KM estimator (that is, without incorporating the rerandomization correction to the variance) will be conservative, in the sense that the resulting CBs are wider than asymptotically necessary. Investigators reporting such estimates should therefore acknowledge that the coverage is achieved with excess interval width, and that the corrected variance estimator derived from our limiting theory should be used to obtain sharper inference. To facilitate adoption, we provide \texttt{R} code implementing the rerandomization-corrected variance estimator for both the KM and IPCW-KM estimators, and the DML estimator at \url{https://github.com/erxc/Surv_rerand}.

The simulation studies underscore the finite-sample reality of these geometric bounds. The magnitude of the variance reduction is bounded by the prognostic power of the rerandomization covariates. In settings where the covariates exhibit weaker correlation with the survival outcome, the geometric deflation factor $\rho_a^\est(t)$ approaches zero. While the exactness of the limiting processes holds regardless of the covariate signal, substantive reductions in the CB ``spindle'' require the pre-specification of highly prognostic baseline metrics. Several extensions to our current results warrant further investigation. First, our theory development assumes a fixed number of rerandomization covariates, $p^\rr$. Extending this geometry to high-dimensional regimes where $p^\rr \to \infty$ as $n \to \infty$ \citep{Wang2022} would require adapting the Mahalanobis distance constraint and establishing uniform Donsker conditions under diverging dimensions. Second, our results focus on standard rerandomization with a fixed set of strata. Extending the current asymptotic framework to finely stratified rerandomization designs \citep{bai2023efficiency, cytrynbaum2024finely} would require establishing uniform weak convergence of survival function estimators under designs where the number of strata grows with sample size, presenting an additional technical challenge from the finite-dimensional M-estimation setting considered in prior works. This will be pursued in future research.

\section*{Acknowledgements}

Research in this article was supported by the United States National Institutes of Health (NIH), National Heart, Lung, and Blood Institute (NHLBI, grant number 1R01HL178513). All statements in this report, including its findings and conclusions, are solely those of the authors and do not necessarily represent the views of the NIH. The authors declare that there are no conflicts of interest relevant to this work.

\singlespacing
\bibliographystyle{jasa3}
\bibliography{surv-rerand}

\end{document}